# AssemblyNet: A Novel Deep Decision-Making Process for Whole Brain MRI Segmentation

Pierrick Coupé[1], Boris Mansencal[1], Michaël Clément[1], Rémi Giraud[2], Baudouin Denis de Senneville[3], Vinh-Thong Ta[1], Vincent Lepetit[1] and José V. Manjon[4]

[1] CNRS, Univ. Bordeaux, Bordeaux INP, LABRI, UMR5800, F-33400 Talence, France
[2] Bordeaux INP, Univ. Bordeaux, CNRS, IMS, UMR 5218, F-33400 Talence, France
[3] CNRS, Univ. Bordeaux, IMB, UMR 5251, F-33400 Talence, France
[4] ITACA, Universitat Politècnica de València, 46022 Valencia, Spain

**Abstract.** Whole brain segmentation using deep learning (DL) is a very challenging task since the number of anatomical labels is very high compared to the number of available training images. To address this problem, previous DL methods proposed to use a global convolution neural network (CNN) or few independent CNNs. In this paper, we present a novel ensemble method based on a large number of CNNs processing different overlapping brain areas. Inspired by parliamentary decision-making systems, we propose a framework called AssemblyNet, made of two "assemblies" of U-Nets. Such a parliamentary system is capable of dealing with complex decisions and reaching a consensus quickly. AssemblyNet introduces sharing of knowledge among neighboring U-Nets, an "amendment" procedure made by the second assembly at higher-resolution to refine the decision taken by the first one, and a final decision obtained by majority voting. When using the same 45 training images, AssemblyNet outperforms global U-Net by 28% in terms of the Dice metric, patch-based joint label fusion by 15% and SLANT-27 by 10%. Finally, AssemblyNet demonstrates high capacity to deal with limited training data to achieve whole brain segmentation in practical training and testing times.

## 1   Introduction

Quantitative brain analysis is crucial to better understand the human brain and to detect pathologies. However, whole brain segmentation is still a very challenging problem, mostly due to the high number of anatomical labels compared to the limited number of available training data. Indeed, manual segmentation of the whole brain is a very tedious and difficult task, preventing the production of large annotated datasets. To address this question, several methods have been proposed in the past years. One of the main references in the domain is the patch-based joint label fusion (JLF) which won the MICCAI challenge in 2012 [1]. More recently, deep leaning (DL) methods have also been proposed. Due to limited GPU memory, first attempts were based on patchwise strategies [2][3] or 2D segmentation (slice by slice) [4]. Last year, first 3D fully convolutional network methods were proposed using reduced input data size (*i.e.*, 128×128×128 voxels) [5] or Spatially Localized Atlas Network Tiles (SLANT) strategy [6]. This latter framework divides the whole volume into overlapping sub-volumes, each one being processed by a different U-Net [7] (*e.g.*, 27). The SLANT strategy addresses the problem of memory and simplifies the complex problem of whole brain segmentation into simpler problems, better suited to limited training data.

In this paper, we propose to extend this framework by using a much larger number of simpler U-Nets (*i.e.*, 250). The main addressed question is the optimal organization of this large ensemble. To this end, we propose a new framework called AssemblyNet. Inspired by decision-making process developed by human societies to deal with complex problems, we decided to model a parliamentary system based on two separate assemblies. Such bicameral – meaning two chambers – parliament has been adopted by many countries around the world. A bicameral system is usually composed of an upper and a lower chamber, both having their own independency to ensure the balance of power. However, an assembly may communicate its vote to the other for amendment. Such parliamentary system is capable of dealing with complex decisions and reaching a consensus quickly.

## 2 Methods

### 2.1 General overview

In AssemblyNet, both assemblies are composed of U-Net considered as "assembly members" (see Fig. 1). Each member represents one territory (*i.e.*, brain area) in the final vote. To this end, we used spatially localized networks where each U-Net only processes a sub-volume of the global volume as done in [6]. Sub-volumes overlap each other, so the final segmentation results from an overcomplete aggregation of local votes. A majority vote is used to obtain the global segmentation. Moreover, each member can share knowledge with his nearest neighbor in the assembly. In particular, we propose a novel nearest neighbor transfer learning strategy, where weights of the spatially nearest U-Net are used to initialize the next U-Net. In addition, we also propose to use prior knowledge on the expected final decision which can be viewed as the bill (*i.e.*, draft law) submitted to an assembly for consideration. As prior knowledge, we decided to use non-linearly registered Atlas priors. Finally, we also propose modeling communication between both assemblies using an innovative strategy. In AssemblyNet, we used a multiscale cascade of assemblies where the first assembly produces coarse decision at 2×2×2 mm. This coarse decision is transmitted to the second assembly for analysis at 1×1×1 mm. This amendment procedure is similar to an error correction or a refinement step. After consideration by both assemblies, the bill under consideration becomes a law which represents the final segmentation in our system.

Our contributions are threefold: *i)* the use of prior knowledge based on fast atlas registration, *ii)* a knowledge sharing between CNNs using nearest neighbor transfer learning and *iii)* an iterative refinement process based on a multiscale cascade of assemblies.

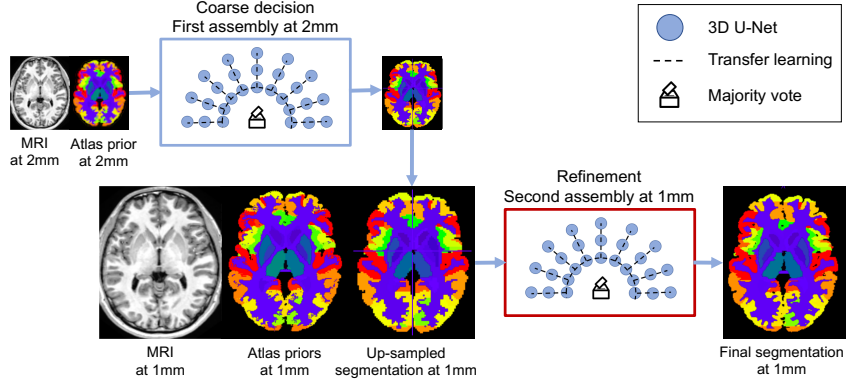

**Fig. 1.** Illustration of the proposed AssemblyNet framework.

### 2.2 Proposed framework

**Preprocessing**: To homogenize input orientations and intensities, all the images are first preprocessed with the following steps: *i)* denoising [8], *ii)* inhomogeneity correction [9], *iii)* affine registration into the MNI space (181×217×181 voxels at 1×1×1 mm) [10], *iv)* tissue-based intensity normalization [11] and *v)* brain extraction [12]. Afterwards, image intensities are centralized and normalized within the brain mask.

**Atlas priors:** To obtain priors knowledge on the expected results, we perform a non-linear registration of the MICCAI 2012 Multi-Atlas Labeling Challenge atlas to the subject under consideration.

**Assembly description:** Each assembly is composed of 125 U-Nets equally distributed in the MNI space along each axis (i.e., 5 along x, y and z). Each 3D U-Net processes a sub-volume large enough to ensure at least 50% of overlap between sub-spaces. At the end, a majority vote is used to aggregate the local votes.

**Nearest neighbor transfer learning:** To enable knowledge sharing between U-Nets of an assembly, we propose a new transfer learning where the weights of the nearest U-Net are used to initialize the next U-Net. In practice, we only copy the weights of the descending path of the U-Net architecture. At the beginning, we train the first U-Net from scratch. Then, each U-Net on the first column is initialized with weights of the previous U-Net. Once the first column is trained, each U-Net of the next column is initialized with the U-Net at the same position on the previous column and so on. Finally, once the first 2D plane of U-Nets is trained, each U-Net of the next 2D planes is initialized with the U-Net at the same position on the previous plane and so on.

**Multiscale cascade of assemblies:** To make our decision-making system faster and more robust, we decided to use a multiscale framework. Consequently, the first assembly at 2×2×2 mm produces a coarse segmentation. Afterwards, an up-sampling of this segmentation to 1×1×1 mm is performed using nearest neighbor interpolation. The second assembly estimates the final result at 1×1×1 mm.

# 3 Experiments

## 3.1 Datasets

**Training dataset**: 45 T1w MRI from the OASIS dataset [13] manually labeled according to the BrainCOLOR protocol were used for training. The selected images were the same than the ones used in [6]. All the used images and manual segmentations are from Neuromorphometrics Inc. During our experiments, we used the 132 anatomical labels consistent across subjects (see [6]).

**Testing dataset**: 19 T1w MRI manually labeled according to the BrainCOLOR protocol were used for testing. These MRI come from three different datasets: 5 from the OASIS dataset, one from the colin27 cohort [14] and 13 from the CANDI database [15]. This testing dataset is the same than the one used in [6].

## 3.2 Implementation details

**Data augmentation**: First, the training images were flipped along mid sagittal plane in the MNI space. Then, we used MixUp data augmentation during training to minimize overfitting problems [16]. This method performs a linear interpolation of a random pair of training examples and their corresponding labels.

**Training framework:** For all the networks, we used the U-Nets architecture proposed in [6], but with a lower number of filters. Instead of using a basis of 32 filters of $3\times3\times3$ – 32 for the first layer, 64 for the second and so on – we selected a basis of 24 filters of $3\times3\times3$ to reduce by 25% the network size. Moreover, we used the same parameters for all the U-Nets with: batch size = 1, optimizer = Adam, epoch = 100, loss = Dice and dropout = 0.5 after each block of the descending path. For the U-Nets of the first assembly at $2\times2\times2$ mm, we used input resolution = $32\times48\times32$ voxels and input channel = 2 (*i.e.*, T1w and Atlas priors). For the U-Nets of the second assembly at $1\times1\times1$ mm, we used input resolution = $64\times72\times64$ voxels and input channel = 3 (*i.e.*, T1w, Atlas priors and up-sampled coarse segmentation). In addition, to compensate for the small batch size, we performed model weights averaging. At the end of the 100 epochs, we performed additional 20 epochs where the model estimated at each epoch is averaged with previous ones. Such average of model weights along the optimization trajectory leads to better generalization than usual training [17]. Finally, we also performed dropout at test time [18]. For each U-Net, we generate 3 different outputs before averaging them. Such method helps reducing variance of the networks. As in [6], the experiments were done with an NVIDIA Titan Xp with 12 GB memory and thus processing times are comparable.

**Computational time:** The preprocessing steps take around 90s. The non-linear registration of the atlas takes less than 5s thanks to a deep leaning framework similar to [19]. The first assembly at $2\times2\times2$ mm requires 3min to segment an image while the second assembly at $1\times1\times1$ mm requires 5min. At the end, the final segmentation is registered back to the native space using inverse affine transform estimated during preprocessing. This interpolation takes around 30s. Therefore, the full AssemblyNet process takes around 10min including preprocessing and inverse registration back to the native space.

### 3.3 Validation framework

First, for each testing subject, we estimated the average Dice coefficient on the 132 considered anatomical labels (without background) in the native space. Afterwards, we estimated the global mean Dice in % over the 19 testing images. In this study, we compared AssemblyNet with several state-of-the-art methods. First, the patch-based joint label fusion (JLF) [1] is used as reference. In addition, we included U-Net [7], SLANT-8 and SLANT-27 methods as proposed in [6]. SLANT-8 is based on 8 U-Nets processing non-overlapping sub-volumes of 86×110×78 voxels while SLANT-27 is based on 27 U-Nets processing overlapping sub-volumes 96×128×88 voxels. All these methods were trained on the 45 training images described in section 3.1. Finally, we included SLANT-27 trained on 5111 auxiliary images segmented using JLF and fined tuned on the 45 training images. This is the best published results for whole brain segmentation to our knowledge. For all these methods we report the results published in [6].

## 4 Results

First, we evaluated the proposed contributions (see Tab. 1). Compared to baseline results at 2×2×2 mm (Dice=67.4%), the used of Atlas priors provided a gain of 0.5% in term of mean Dice. Moreover, the combination of Atlas priors and transfer learning improved by 0.7% the baseline mean Dice. In addition, multiscale cascade of assemblies increased by 1.6% the mean Dice obtained with Assembly at 1×1×1 mm without multiscale cascade (Dice=72.2%). Finally, AssemblyNet outperformed by 8.7% the mean Dice obtained with baseline Assembly at 2×2×2 mm.

**Table 1.** Evaluation of the proposed contributions. The mean Dice is evaluated on the 19 images of the test dataset in the native space for the 132 considered labels (without background). Testing time includes image preprocessing and registration back to the native space.

| Methods | Atlas prior | Transfer learning | Multi-scale | Mean Dice in % | Training time | Testing time |
|---|---|---|---|---|---|---|
| Assembly at 2×2×2 mm | No | No | - | 67.4 | 29h | 5min |
| Assembly at 2×2×2 mm | Yes | No | - | 67.7 | 29h | 5min |
| Assembly at 2×2×2 mm | Yes | Yes | - | 67.9 | 29h | 5min |
| Assembly at 1×1×1 mm | Yes | Yes | No | 72.2 | 6 days | 7min |
| AssemblyNet | Yes | Yes | Yes | **73.3** | 7 days | 10min |

Afterwards, we compared AssemblyNet with state-of-the-art methods (see Tab. 2). When considering only methods trained with 45 images, AssemblyNet improved mean Dice obtained with U-Net and SLANT-8 by 28%, JLF by 15% and SLANT-27 by 10%. AssemblyNet was also efficient in term of training and testing times compared to SLANT-based methods. It has to be noted that Assembly at 2×2×2 mm outperformed all the methods except AssemblyNet while working at lower resolution. Finally, compared to SLANT-27 trained over 5111+45 images, our method provided slightly better results (without library extension) while being faster to train and to execute. According to [6], their library extension required 21 CPU years to be completed. Consequently, such an approach is impractical or very costly using a cloud-based solution. Therefore, all these results demonstrate that AssemblyNet is very efficient to deal with limited training data and to accurately achieve segmentation in practical training and testing times.

**Table 2.** Methods comparison on the 19 images of the testing dataset. The mean Dice is evaluated on the 132 considered labels (without background) in the native space.

| Methods | Number of training images | Mean Dice in % | Training time | Testing time | Library extension time |
|---|---|---|---|---|---|
| U-Net [6] | 45 | 57.0 | 33h | 8min | 0s |
| SLANT-8 [6] | 45 | 57.0 | 11 days | 10min | 0s |
| JLF [1] | 45 | 63.4 | 0s | 34h | 0s |
| SLANT-27 [6] | 45 | 66.1 | 42 days | 15min | 0s |
| Assembly at 2×2×2 mm | 45 | 67.9 | 29h | 5min | 0s |
| SLANT-27 [6] | 5111 + 45 | 72.9 | 27 days | 15min | 21 years* |
| AssemblyNet | 45 | **73.3** | 7 days | 10min | 0s |

\* Library extension time represents the CPU time required to segment 5111 MRI using JLF (*i.e.*, 34h*5111). This number of 21 CPU years is reported in [6].

Finally, we analyzed the performance of the methods according to the dataset. Mean Dice coefficients obtained on each testing dataset (*i.e.*, OASIS, CANDI and Colin27) are provided in Tab. 3. As expected, all the methods performed better on adult scans from the OASIS dataset since the training dataset comes from the same cohort. On this dataset AssemblyNet outperformed all the methods. On child scans from the CANDI dataset acquired with different protocols, we can note a dramatic drop in performance for all the methods except for AssemblyNet and SLANT-27 trained on 5111+45 images. The auxiliary library used by SLANT-27 is based on 9 different databases (64 acquisition sites) and includes more than 1000 scans of child making this method able to segment child MRI. By using only 45 adult scans coming from a single acquisition site, AssemblyNet produced similar results on this dataset. This demonstrates the robustness of the proposed framework to unseen acquisition protocol and ages. Finally, on the high-resolution Colin27 image, AssemblyNet obtained the best segmentation accuracy. As for SLANT-27, one could expect segmentation improvements for AssemblyNet using library extension. We will investigate such framework in future work.

**Table 3.** Methods comparison on the different testing datasets (5 adult scans from OASIS, 13 child scans from CANDI child and the high-resolution Colin27 image based on scans average).

| Methods | Number of training images | OASIS Mean Dice in % | CANDI Mean Dice in % | Colin27 Dice in % |
|---|---|---|---|---|
| U-Net [6] | 45 | 70.6 | 51.4 | 62.1 |
| SLANT-8 [6] | 45 | 69.9 | 51.9 | 59.7 |
| JLF [1] | 45 | 74.6 | 59.0 | 64.6 |
| SLANT-27 [6] | 45 | 76.6 | 62.1 | 66.5 |
| SLANT-27 [6] | 5111 + 45 | 77.6 | **71.1** | 73.2 |
| AssemblyNet | 45 | **78.8** | **71.1** | **74.2** |

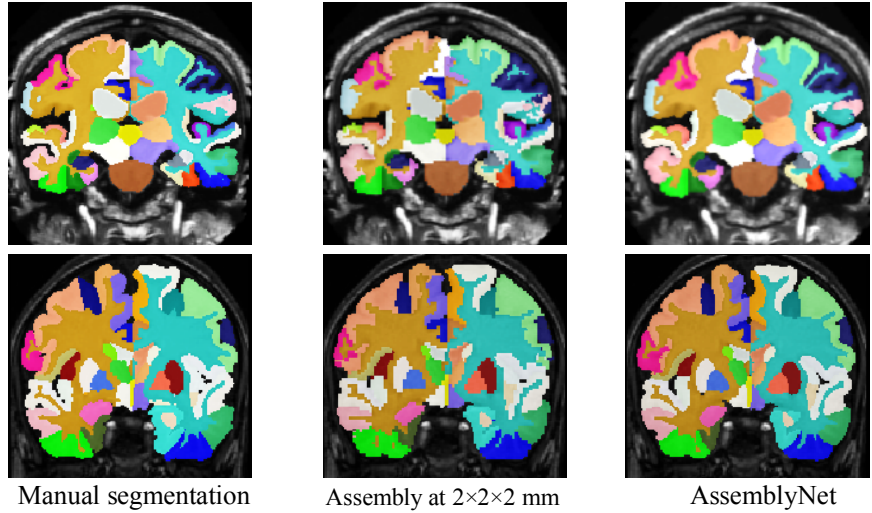

| Manual segmentation | Assembly at 2×2×2 mm | AssemblyNet |

**Fig. 2.** Examples of segmentation in the native space obtained for one test subject from OASIS (at the top) and one test subject from CANDI (at the bottom). Note the accuracy of the segmentation compared to the manual segmentation.

Fig. 1 shows segmentation in the native space for a subject from the OASIS dataset and one for the CANDI cohort. This figure shows segmentation at 2×2×2 mm and the improvement obtained using refinement at 1×1×1 mm with the second assembly.

## 5   Conclusion

In this paper, we proposed to use of a large number of CNNs to perform whole brain segmentation. We investigated how to organize this large ensemble of CNNs to quickly and accurately segment the brain. To this end, we designed a novel deep decision-making process called AssemblyNet based on two assemblies of U-Nets. Our validation showed the very competitive results of AssemblyNet compared to state-of-the-art methods. We also demonstrated that AssemblyNet is very efficient to deal with limited training data and to accurately achieve segmentation in a practical training and testing times.

**Acknowledgement**: This work benefited from the support of the project DeepvolBrain of the French National Research Agency (ANR-18-CE45-0013). This study was achieved within the context of the Laboratory of Excellence TRAIL ANR-10-LABX-57 for the BigDataBrain project. Moreover, we thank the Investments for the future Program IdEx Bordeaux (ANR-10-IDEX- 03- 02, HL-MRI Project), Cluster of excellence CPU and the CNRS. This study has been also supported by the DPI2017-87743-R grant from the Spanish Ministerio de Economia, Industria Competitividad. The authors gratefully acknowledge the support of NVIDIA Corporation with their donation of the TITAN Xp GPU used in this research.